\documentclass[12pt]{article}
\usepackage{graphicx}

\title{Kondo Resonance Decoherence by an External
Potential 
}
\author{R. C. Monreal and F. Flores \\ 
Departamento de F\'{\i}sica Te\'orica de la Materia
Condensada, C-V . \\
Universidad Aut\'onoma de Madrid, E-28049 Madrid,
Spain. \\ }

\begin {document}

\maketitle

\begin{abstract}
The Kondo problem, for a quantum dot (QD), subjected to
 an external
bias, is analyzed in the limit of infinite Coulomb repulsion
by using a consistent equations of motion
method based on a slave-boson Hamiltonian. Utilizing
a strict perturbative solution in the leads-dot
coupling, $T$, up to $T^4$ and $T^6$ orders, we
calculate the QD spectral density and conductance, as
well as the decoherent rate that drive the system from
the strong to the weak coupling regime. Our results
indicate that the weak coupling regime is reached for
voltages larger than a few units of the Kondo
temperature.
 
\end{abstract}

PACS: 72.15.Qm, 73.23.Hk, 75.20.Hr 

\newpage

\section{Introduction}
The transport of electrons through a quantum dot (QD)
is a subject of increasing interest for its many
applications in nanotechnology. 
 Although there exists exact solutions for the
ground state of the QD connected to metallic leads \cite{Hew}, 
no such methods are applicable to 
nonequilibrium systems. Typical cases are 
those of (a) a bias or (b) an electromagnetic radiation between
the leads of a quantum well \cite{Ka}.  It has been suggested
that, in case (a), two Kondo resonances appear, each one
associated with every lead \cite{WM1},\cite{Jap}.
 Some experimental evidence has
been presented supporting this view \cite{Exp}. 
Still, it is not
well understood how inelastic (decoherent) processes
modify the strong-coupling regime associated with the
Kondo resonance. 
 To our knowledge, there are only a few
theoretical approaches to the problem.
The non-crossing
approximation (NCA) \cite{WM1},\cite{WM3}, being 
essentially numerical, can give analytical results only
in the limit $Ln(V/T_{K})>>1$ (V being the applied bias
and $T_{K}$ the Kondo energy), the same limit in
which renormalized
perturbation theory (RPT) \cite{RPT},\cite{RPT1} is valid.
 It is
important, however, to have analytical 
results for the decoherence rate, not only
to make theoretical predictions, but also to be able to
interpret correctly the results of calculations done
for more complicated situations, i.e., when time-
dependent potentials are applied to the QD
\cite{Pli}-\cite{Jaime}.
A version of the nonequilibrium Kondo problem has been
solved exactly to all orders in the applied bias V in 
\cite{SS}.

In this article, we address the nonequilibrium
 Kondo problem for a QD
in the limit of infinite Coulomb repulsion U, 
using an equation of motion (EOM)
method  based on a slave-boson Hamiltonian. By means of
a strict perturbative solution in the leads-dot
coupling, $T$, to orders $T^4$ and $T^6$, we have been
able to identify the decoherent processes that broaden
the Kondo-level, as a function of the applied
bias, and break the strong-coupling regime. 
The EOM 
method has been applied before 
to the Kondo problem
\cite{Lacroix} but it was mainly used 
with approximations valid only for high
temperatures \cite{WM1},\cite{WM2},\cite{Pol}.
 We show that this
method, when applied consistently to high orders,
can be conveniently used in the nonequilibrium
case, describing well the inelastic processes
contributing to the Kondo resonance decoherence.
We first check the method in the equilibrium regime,
comparing our results with 
exact results using the numerical
renormalization group (NRG) and also with the
NCA.  When used in the nonequilibrium case, we find
that, for an applied bias larger than $2T_K$, the system 
develops 
two-peaked resonant structures at
$\pm V/2$; we show, however, that decoherent processes  
take the system from the strong-coupling regime, where
the spin of the impurity is quenched by the lead
electrons, 
to the weak-coupling regime, if 
$V/T_K$ is larger than a few units.
We conclude that our method
provides an accurate and powerful tool to
analyze nonequilibrium systems and time-dependent
problems,  situations where
very few theoretical approaches are valid.

\section{Theory}
Our starting point for a QD level with
infinite-U, coupled to two leads, is the 
slave-boson Hamiltonian for
spin up and down electrons \cite{SB}:

\begin{eqnarray}
\hat{H}=\sum_{k \sigma}\epsilon_{k \sigma}
\hat{c}_{k \sigma}^{\dagger}
\hat{c}_{k\sigma} +\sum_{\sigma} \epsilon_0
\hat{d}_{\sigma}^{\dagger}\hat{d}_{\sigma} +
\sum_{k \sigma} T_k (
\hat{c}_{k \sigma}^{\dagger} \hat{\chi}_{\sigma}+
\hat{\chi}_{\sigma}^{\dagger} \hat{c}_{k \sigma})
\label{Ham}
\end{eqnarray}
where the operators $\hat{\chi}_{\sigma}$ and
$\hat{\chi}_{\sigma}^{\dagger}$ are
introduced in terms of an auxiliary boson field, 
$\hat{b}$ and $\hat{b}^{\dagger}$, as 
$\hat{\chi}_{\sigma}=\hat{b}^{\dagger} \hat{d}_{
\sigma}$,   
$\hat{\chi}_{\sigma}^{\dagger}=\hat{d}_{\sigma}^{\dagger}
\hat{b}$
and $k={k_L,k_R}$ for left and right leads.
 In eq.(\ref{Ham}), the operator $\hat{c}_{k
\sigma}^{\dagger}$ ($\hat{c}_{k \sigma}$) creates 
(anhilates) an electron of momentum $k$ and spin
$\sigma$ in one of the leads, while $\hat{d}_{\sigma}
^{\dagger}$  ($\hat{d}_{\sigma}$) creates 
(anhilates) an electron of spin $\sigma$ in the dot.
The constraint relation

\begin{equation}
\hat{b}^{\dagger}\hat{b}+\sum_{\sigma}
\hat{d}_{\sigma}^{\dagger}\hat{d}_{\sigma}=1
\label{constraint}
\end{equation}
is imposed to guarantee that no more than one 
electron occupies the QD since we are in the
infinite-U limit.

The physical (retarded) Green function for this problem
is defined as \cite{Bick},\cite{T2}
\begin{equation}
<<\hat{\chi}_{\sigma}(t);\hat{\chi}_{\sigma}^{\dagger}(t')>>=
-i<\hat{\chi}_{\sigma}^{\dagger}(t')\hat{\chi}_{\sigma}(t)+
\hat{\chi}_{\sigma}(t) \hat{\chi}_{\sigma}^{\dagger}(t')>
\theta(t-t')
\label{chi}
\end{equation}
and its Fourier transform is written as 
$<<\hat{\chi}_{\sigma};\hat{\chi}_{\sigma}^{\dagger}>>
(\omega)$.  The EOM for this Green
function reads (omitting the argument $\omega$ in all
the Green functions):

\begin{eqnarray}
(w- \epsilon_0)
<<\hat{\chi}_{\sigma};\hat{\chi}_{\sigma}^{\dagger}>>=
1-<\hat{n}_{-\sigma}>+  
\sum_{k}T_k <<(1-\hat{n}_{-\sigma})\hat{c}_{k
\sigma};\hat{\chi}_{\sigma}^{\dagger}>>+
\nonumber \\
\sum_{k}T_k <<\hat{d}_{-\sigma}^{\dagger}
\hat{d}_{\sigma}\hat{c}_{k -\sigma};
\hat{\chi}_{\sigma}^{\dagger}>>
\label{chiw}
\end{eqnarray}
where the new Green functions appearing on the
right-hand side of eq.(\ref{chiw}) are defined 
as in eq.(\ref{chi}). $\hat{n}_{-\sigma}$ is the number
operator for electrons of spin $-\sigma$ in the QD and
$<\hat{A}>$ will denote the expectation value of
 $\hat{A}$ from now on. 
In a second step, the EOM for these Green
functions lead to a second generation of Green
functions, viz.,  

\begin{equation}
(w-\epsilon_k)
<<\hat{c}_{k\sigma};\hat{\chi}_{\sigma}^{\dagger}>>=
T_k <<\hat{\chi}_{\sigma};\hat{\chi}_{\sigma}^{\dagger}>>
\label{2a}
\end{equation}

\begin{eqnarray}
(w-\epsilon_k)
<<\hat{d}_{-\sigma}^{\dagger}
\hat{d}_{\sigma}\hat{c}_{k -\sigma}
;\hat{\chi}_{\sigma}^{\dagger}>>=
-<\hat{\chi}_{-\sigma}^{\dagger} \hat{c}_{k -\sigma}>+
\nonumber \\
\sum_{k'}T_{k'}
<<\hat{\chi}_{\sigma} \hat{c}_{k' -\sigma}^{\dagger}
\hat{c}_{k -\sigma}
;\hat{\chi}_{\sigma}^{\dagger}>>+
\sum_{k'}T_{k'}
<<\hat{\chi}_{-\sigma}^{\dagger}\hat{c}_{k -\sigma}
\hat{c}_{k' \sigma}
;\hat{\chi}_{\sigma}^{\dagger}>>
\label{2b}
\end{eqnarray}

\begin{eqnarray}
(w-\epsilon_k)
<<\hat{n}_{-\sigma} \hat{c}_{k\sigma}
;\hat{\chi}_{\sigma}^{\dagger}>>=
\sum_{k'}T_{k'}
<<\hat{\chi}_{-\sigma} \hat{c}_{k' -\sigma}^{\dagger} 
\hat{c}_{k \sigma}
;\hat{\chi}_{\sigma}^{\dagger}>>+
\nonumber \\
\sum_{k'}T_{k'}
<<\hat{\chi}_{-\sigma}^{\dagger} \hat{c}_{k' -\sigma}
\hat{c}_{k \sigma}
;\hat{\chi}_{\sigma}^{\dagger}>>
\label{2c}
\end{eqnarray}

Note that the second generation of Green functions
involve the excitation of two electrons in the leads. In
our approximation to $O(T^2)$, we contract operators 
$\hat{c}_{k \sigma}$, $\hat{c}_{k' \sigma}^{\dagger}$,
where possible and obtain a closed equation for
$<<\hat{\chi}_{\sigma};\hat{\chi}_{\sigma}^{\dagger}>>$
. In particular we take
$<<\hat{\chi}_{\sigma} \hat{c}_{k' -\sigma}^{\dagger}
\hat{c}_{k -\sigma}
;\hat{\chi}_{\sigma}^{\dagger}>>=\delta_{k,k'}
<\hat{n}_{k -\sigma}>
<<\hat{\chi}_{\sigma};\hat{\chi}_{\sigma}^{\dagger}>>$
and
$
<<\hat{\chi}_{-\sigma}^{\dagger}\hat{c}_{k -\sigma}
\hat{c}_{k' \sigma}
;\hat{\chi}_{\sigma}^{\dagger}>>=
<<\hat{\chi}_{-\sigma} \hat{c}_{k' -\sigma}^{\dagger} 
\hat{c}_{k \sigma}
;\hat{\chi}_{\sigma}^{\dagger}>>=
<<\hat{\chi}_{-\sigma}^{\dagger} \hat{c}_{k' -\sigma}
\hat{c}_{k \sigma}
;\hat{\chi}_{\sigma}^{\dagger}>>=0$.
This yields the following approximation to
$<<\hat{\chi}_{\sigma};\hat{\chi}_{\sigma}^{\dagger}>>$,
which is exact 
to  $O(T^2)$ (see Ref. \cite{T2}):

\begin{equation}
<<\hat{\chi}_{\sigma};\hat{\chi}_{\sigma}^{\dagger}>>=
{1-<\hat{n}_{-\sigma}> -\sum_{k}
{T_k
<\hat{\chi}_{-\sigma}^{\dagger} \hat{c}_{k -\sigma}>
\over \omega-\epsilon_k}
\over
\omega-\epsilon_0-\Sigma_0-\sum_k{T_k^2
<\hat{n}_k> \over
\omega-\epsilon_k}}
\label{V2}
\end{equation}
where

\begin{equation}
\Sigma_0=\sum_{k} {T_k^2 \over \omega-\epsilon_k}
\end{equation}

We can obtain 
$<<\hat{\chi}_{\sigma};\hat{\chi}_{\sigma}^{\dagger}>>$  
to $O(T^4)$  by making
all possible contractions of operators 
$\hat{c}_{k \sigma}$, $\hat{c}_{k' \sigma}^{\dagger}$
in a third generation of Green functions.
For instance, the EOM for 
$<<\hat{\chi}_{-\sigma}^{\dagger} \hat{c}_{k' -\sigma} 
\hat{c}_{k \sigma}
;\hat{\chi}_{\sigma}^{\dagger}>>$ in eq.(\ref{2c}) is:
%
%
%

\begin{eqnarray}
(\omega+\epsilon_0-\epsilon_{k}-\epsilon_{k'})
<<\hat{\chi}_{-\sigma}^{\dagger} \hat{c}_{k' -\sigma}
\hat{c}_{k \sigma}
;\hat{\chi}_{\sigma}^{\dagger}>>=
T_{k'}
<<\hat{n}_{-\sigma}\hat{c}_{k \sigma}
;\hat{\chi}_{\sigma}^{\dagger}>>-
\nonumber \\
T_{k'}
<<\hat{d}_{-\sigma}^{\dagger} \hat{d}_{\sigma}
\hat{c}_{k' -\sigma}
;\hat{\chi}_{\sigma}^{\dagger}>>+
\sum_{k''} T_{k''}
<<\hat{d}_{-\sigma}^{\dagger} \hat{d}_{\sigma}
\hat{c}_{k' -\sigma}
\hat{c}_{k'' \sigma}^{\dagger} \hat{c}_{k \sigma}
;\hat{\chi}_{\sigma}^{\dagger}>>-
\nonumber \\
\sum_{k''} T_{k''}
<<(1-\hat{n}_{\sigma}) \hat{c}_{k'' -\sigma}^{\dagger}
\hat{c}_{k' -\sigma} \hat{c}_{k \sigma}
;\hat{\chi}_{\sigma}^{\dagger}>>
\end{eqnarray}
which is approximated by

\begin{eqnarray}
(\omega+\epsilon_0-\epsilon_{k}-\epsilon_{k'})
<<\hat{\chi}_{-\sigma}^{\dagger} \hat{c}_{k' -\sigma}
\hat{c}_{k \sigma}
;\hat{\chi}_{\sigma}^{\dagger}>>=
T_{k'}
<<\hat{n}_{-\sigma}\hat{c}_{k \sigma}
;\hat{\chi}_{\sigma}^{\dagger}>>- 
\nonumber \\
T_{k'}
<<\hat{d}_{-\sigma}^{\dagger} \hat{d}_{\sigma}
\hat{c}_{k' -\sigma}
;\hat{\chi}_{\sigma}^{\dagger}>>+
T_{k} <\hat{n}_{k}>
<<\hat{d}_{-\sigma}^{\dagger} \hat{d}_{\sigma}
\hat{c}_{k' -\sigma}
;\hat{\chi}_{\sigma}^{\dagger}>>- \nonumber \\
T_{k'} <\hat{n}_{k'}>
<<(1-\hat{n}_{\sigma})\hat{c}_{k \sigma}
;\hat{\chi}_{\sigma}^{\dagger}>>
\end{eqnarray}
This procedure yields  the Green functions already 
obtained in eqs.(\ref{2a}-\ref{2c}) and a new one,
$<<\hat{n}_{\sigma}\hat{c}_{k \sigma}
;\hat{\chi}_{\sigma}^{\dagger}>>$, 
which has to be calculated as in eq.(\ref{2c}).  
The first generation of Green functions are thus
obtained as:

\begin{eqnarray}
(\tilde\omega_k-\Sigma_1(\tilde\omega_k))
<<\hat{d}_{-\sigma}^{\dagger} \hat{d}_{\sigma}
\hat{c}_{k -\sigma}
;\hat{\chi}_{\sigma}^{\dagger}>>=
-<\hat{\chi}_{-\sigma}^{\dagger} \hat{c}_{k -\sigma}>+
\nonumber \\
T_k <n_k>
<<\hat{\chi}_{\sigma};\hat{\chi}_{\sigma}^{\dagger}>>-
T_k <\hat{n}_k> \sum_{k'} {T_{k'}
<<\hat{n}_{\sigma} \hat{c}_{k' \sigma}
;\hat{\chi}_{\sigma}^{\dagger}>> \over
\tilde\omega_k -\epsilon_{k'}+\epsilon_0}
\label{3a}
\end{eqnarray}

\begin{eqnarray}
(\tilde\omega_k-\Sigma_2(\tilde\omega_k))
<<\hat{n}_{-\sigma} \hat{c}_{k \sigma}
;\hat{\chi}_{\sigma}^{\dagger}>>=
\Sigma_3(\tilde\omega_k)
<<\hat{n}_{\sigma} \hat{c}_{k \sigma}
;\hat{\chi}_{\sigma}^{\dagger}>> +
\nonumber \\
T_k <\hat{n}_k> \sum_{k'}{T_{k'}
<<\hat{d}_{-\sigma}^{\dagger} \hat{d}_{\sigma}
\hat{c}_{k' -\sigma}
;\hat{\chi}_{\sigma}^{\dagger}>> \over
\tilde\omega_k-\epsilon_{k'}+\epsilon_0}
\label{3b}
\end{eqnarray}

\begin{eqnarray}
(\tilde\omega_k-\Sigma_2(\tilde\omega_k))
<<\hat{n}_{\sigma} \hat{c}_{k \sigma}
;\hat{\chi}_{\sigma}^{\dagger}>>=
-<\hat{\chi}_{\sigma}^{\dagger} \hat{c}_{k \sigma}>+
T_{k} <\hat{n}_{k}>
<<\hat{\chi}_{\sigma};\hat{\chi}_{\sigma}^{\dagger}>>+
\nonumber \\
\Sigma_3(\tilde\omega_k) 
<<\hat{n}_{-\sigma} \hat{c}_{k \sigma}
;\hat{\chi}_{\sigma}^{\dagger}>> -
T_{k} <\hat{n}_{k}> \sum_{k'}{T_{k'}
<<\hat{n}_{-\sigma} \hat{c}_{k' \sigma}
;\hat{\chi}_{\sigma}^{\dagger}>> \over
\tilde\omega_k-\epsilon_{k'}+\epsilon_0}
\label{3c}
\end{eqnarray}
with
$\tilde\omega_k=\omega-\epsilon_{k}$ from now on.
Eqs.(\ref{3a}-\ref{3c})
 define a set of coupled integral equations in the
energy levels of the leads, $\epsilon_k$, whose solution 
yields the Green functions to be introduced in
eq.(\ref{chiw}). 
In these equations we have neglected expectation values
of operators such as
$<\hat{\chi}_{\sigma}^{\dagger}\hat{\chi}_{-\sigma}
\hat{c}_{k -\sigma}^{\dagger}\hat{c}_{k \sigma}>$,
which give small contributions to
$<<\hat{\chi}_{\sigma};\hat{\chi}_{\sigma}^{\dagger}>>$
to $O(T^4)$ and have 
defined several
self-energies given by:

\begin{equation}
\Sigma_i(\tilde\omega_k)=\sum_{k'}T_{k'}^2 \nu_i
[{1 \over
\tilde\omega_k-\epsilon_{k'}+\epsilon_0}+
{1 \over
\tilde\omega_k+\epsilon_{k'}-\epsilon_0}]
\label{sigmai}
\end{equation}
where $\nu_1=1-<\hat{n}_{k'}>$, $\nu_2=1$ and 
$\nu_3=<\hat{n}_{k'}>$,
$(\Sigma_3(\tilde\omega_k)=\Sigma_2(\tilde\omega_k)-
\Sigma_1(\tilde\omega_k)$). 
 These self-energies are associated with different Green
functions and play a crutial role in the Kondo-resonance
decoherence, as we will see below. 
Eqs.(\ref{3a}-\ref{3c}) reproduce the conventional 
$T^2$-approximation of \cite{T2} by neglecting
all the terms proportional to $T^3$.
One can get, however, a new (and effective) $T^2$-order
near the Kondo resonance by 
neglecting the terms that couple these equations in
an integral way.
 This approximation yields the
following Green function:

\begin{eqnarray}
&&<<\hat{\chi}_{\sigma};\hat{\chi}_{\sigma}^{\dagger}>>=
\nonumber \\
&& [1-<\hat{n}_{-\sigma}> -\sum_k 
{T_k
<\hat{\chi}_{-\sigma}^{\dagger} \hat{c}_{k -\sigma}>
\over \tilde\omega_k-\Sigma_1(\tilde\omega_k)}+\sum_k T_k
{\Sigma_3(\tilde\omega_k) 
<\hat{\chi}_{\sigma}^{\dagger} \hat{c}_{k \sigma}>
\over
(\tilde\omega_k-\Sigma_2(\tilde\omega_k))
(\tilde\omega_k-\Sigma_4(\tilde\omega_k))}]
\nonumber \\ 
&&[\omega-\epsilon_0-\Sigma_0-\sum_k{T_k^2
<\hat{n}_k> \over
\tilde\omega_k-\Sigma_1(\tilde\omega_k)}+
\sum_k{T_k^2 <\hat{n}_k> \Sigma_3(\tilde\omega_k) \over
(\tilde\omega_k-\Sigma_2(\tilde\omega_k))
(\tilde\omega_k-\Sigma_4(\tilde\omega_k))}]^{-1}
\label{V4}
\end{eqnarray}
with

\begin{equation}
\Sigma_4(\tilde\omega_k)=\Sigma_2(\tilde\omega_k)+
{\Sigma_3(\tilde\omega_k)^2 \over
\tilde\omega_k-\Sigma_2(\tilde\omega_k)}. 
\end{equation}
In eq.(\ref{V4}), $<n_{-\sigma}>$ and
$<\hat{\chi}_{-\sigma}^{\dagger} \hat{c}_{k -\sigma}>$
have to be calculated self-consistently 
\cite{Lacroix},\cite{T2}. At zero temperature and in a
wide-band approximation, they are
given by:

\begin{equation}
<\hat{n}_{-\sigma}>=\int_{-\infty}^{0}
-{d\omega \over \pi}
Im<<\hat{\chi}_{\sigma};\hat{\chi}_{\sigma}^{\dagger}>>
(\omega)
\end{equation}
and
\begin{equation}
<\hat{\chi}_{-\sigma}^{\dagger} \hat{c}_{k -\sigma}>=
\int_{-\infty}^{0}-{d\omega \over \pi}
Im {T_{k}
<<\hat{\chi}_{\sigma};\hat{\chi}_{\sigma}^{\dagger}>>
(\omega) \over \tilde\omega_{k}}
\end{equation}
where $\omega=0$ is the Fermi level.

\section{Results}
In this section we first discuss some consequences that 
our approximation bears in the description of the Kondo
peak. Notice that the $T^2$-order of eq.(\ref{V2})
is given by eq.(\ref{V4})
neglecting the last terms of the numerator and
denominator and taking $\Sigma_1(\tilde\omega_k)=0$. 
These new terms
yield, however, contributions that are proportional to
$T^2$ around the Kondo peak. This can be seen 
by considering  the
self-energies $\Sigma_1$, $\Sigma_2$ and $\Sigma_4$,
whose imaginary parts tend to broaden the Kondo-like
peak associated with the factors 
$\sum_{k} {T_k^2 <\hat{n}_k> \over
\tilde\omega_k-\Sigma_i(\tilde\omega_k)}$.
In the $T^4$-order, eq.(\ref{sigmai}) yields, 
for $V=0$, $\epsilon_0<0$, 
$|\epsilon_0|>>\Delta$ ($2\Delta$ being the one-electron
level-width) 
and a wide-band
approximation,
$\Sigma_2=2 \Sigma_0\equiv-2i\Delta$, $\Sigma_3 \simeq
\Sigma_2$  and
$\Sigma_1=0$, 
if $|\omega-\epsilon_k|<<|\epsilon_0|$.
Then, near the Fermi level, we obtain:
\begin{equation}
\sum_{k} {T_k^2 <\hat{n}_k> \over \tilde\omega_k-
\Sigma_1(\tilde\omega_k)} \simeq {\Delta \over \pi}Ln({D \over \omega})
\end{equation}
 D being the half-bandwidth,  and

\begin{equation}
\sum_{k} {T_k^2 <\hat{n}_k> \over
\tilde\omega_k-\Sigma_2(\tilde\omega_k)}
{\Sigma_3(\tilde\omega_k) \over
\tilde\omega_k-\Sigma_4(\tilde\omega_k)} \simeq
-{\Delta \over 2\pi}Ln({\omega+4i\Delta \over
\omega}).
\label{new}
\end{equation}
The important point is that the term given by
eq.(\ref{new}), which
is neglected in the $T^2$-order approximation, also
gives a $T^2$-contribution to
$<<\hat{\chi}_{\sigma};\hat{\chi}_{\sigma}^{\dagger}>>$
going like $Ln(\omega)$ at the Fermi energy.
In particular, eq.(\ref{V4}) yields the Kondo-energy 

\begin{equation}
T_K\simeq D^{2/3} (4\Delta)^{1/3} 
exp[{-2\pi |\epsilon_0| \over 3\Delta}]
\label{wkondo}
\end{equation}
intermediate between 
$D exp[-\pi |\epsilon_0|/\Delta]$
(the $T^2$-order) and the exact result of Haldane 
$(D\Delta)^{1/2}exp[-\pi |\epsilon_0|/2 \Delta]$
\cite{Hal}.

A check of our solution, eq.(\ref{V4}), can be obtained
 by considering
the case $V=0$; this equilibrium case can be compared
with the NCA and with the numerically exact calculations 
of Ref.\cite{Costi} using the NRG.
Fig.1 shows our calculated spectral density  for
the Kondo regime defined by $\epsilon_0=-4\Delta$,
$\Delta=0.01$ and $D=1$ at zero temperature. 
Two cases, $T^2$-order and
eq.(\ref{V4}) are shown as a function of $\omega/T_K$. 
Although for this particular
case our value of $T_K$, eq.(\ref{wkondo}), is very similar to Haldane's,
we use this last one from now on.
Our results indicate that eq.(\ref{V4}) yields a
much better spectral density for the Kondo-resonance than the
$T^2$-order aproximation, with values very similar to
the ones given by the NRG above the Fermi energy, 
although our results yield a
too small value of the spectral density below the Fermi
energy; this deviation extends to energies of around
$2T_K$ and suggests that a much more accurate solution
is going to be obtained for voltages larger than this
value (see below).
Fig.2 shows the same results for a mixed-valence
regime ($\epsilon_0=0$,
$\Delta=0.01$ and $D=1$) at zero temperature. In this case,
eq.(\ref{V4}) yields an excellent approximation to
the exact NRG-calculation and the Friedel-Langreth
sum rule is satisfied with an accuracy better than
10$\%$. The agreement is even better 
in the empty orbital regime, Fig.3, where the sum rule is
fulfilled to within 5$\%$ accuracy. 
We should mention that our present results are in much better
agreement with the exact results than
the ones obtained using
Lacroix's approximation \cite{Lacroix}
 and shown in \cite{T2}.

Now we turn our attention to the nonequilibrium case
$V\neq 0$ in the symmetric situation, where the
chemical potentials of the left and right leads are
 $\pm V/2$. A fully consistent solution of the
case $V\neq 0$ requires the application of the Keldysh
formalism, as explained in \cite{T2}. The symmetric case
can be analyzed, however, at the level of the above
equations because, to good accuracy,  
$<n_{-\sigma}>$ and
$<\hat{\chi}_{-\sigma}^{\dagger} \hat{c}_{k -\sigma}>$
do not vary with V for the values used here.
 We start by analyzing the
self-energies
$\Sigma_1$ and $\Sigma_4$  in the regime
$\tilde\omega_k \rightarrow 0$,
which is the relevant regime for analyzing the
Kondo-resonance decoherence. It is easy to see 
that eq.(\ref{sigmai}) produces 
$\Sigma_1(\tilde\omega_k \rightarrow 0)\rightarrow 0$ 
, as in the equilibrium case and therefore, 
we find no decoherence for $V\neq 0$. 
It is interesting to proceed further with the EOM method
 and see how these self-energies are
modified. 
By going to $O(T^6)$, we find for $\Sigma_1$ and
$\Sigma_2$:

\begin{eqnarray}
\Sigma_1(\tilde\omega_k)=\sum_{k'}T_{k'}^2
[{1-<\hat n_{k'}> \over \tilde\omega_k-\epsilon_{k'}+
\epsilon_{-}(k')+i\Delta}+
{1-<\hat n_{k'}> \over \tilde\omega_k+\epsilon_{k'}-
\epsilon_{+}(k')+i\Delta}]+ 
\nonumber \\
2 \sum_{k',k''} {T_{k'}^2 (1-<\hat n_{k'}>) \over
\tilde\omega_k-\epsilon_{k'}+
\epsilon_{-}(k')+i\Delta}
{1 \over
\tilde\omega_k+\epsilon_{k''}-\epsilon_{k'}}
{T_{k''}^2 (1-<\hat n_{k''}>) \over
\tilde\omega_k+\epsilon_{k''}-
\epsilon_{+}(k'')+i\Delta}+ \nonumber \\
\sum_{k',k'',q} {T_{k'}^2 (1-<\hat n_{k'}>) \over
\tilde\omega_k-\epsilon_{k'}+
\epsilon_{-}(k')+i\Delta}
{1 \over
\tilde\omega_k+\epsilon_{k''}-\epsilon_{k'}}
{T_{k''}^2 (1-<\hat n_{k''}>) \over
\tilde\omega_k+\epsilon_{k''}-
\epsilon_{+}(k'')+i\Delta}\times
\nonumber \\
{1 \over
\tilde\omega_k+\epsilon_{k''}-\epsilon_{q}}
{T_{q}^2 (1-<\hat n_{q}>) \over
\tilde\omega_k-\epsilon_{q}+
\epsilon_{-}(q)+i\Delta} + O(T^8)
\label{sigma1-4}
\end{eqnarray}
and
\begin{eqnarray}
\Sigma_2(\tilde\omega_k)=\sum_{k'}T_{k'}^2
[{1 \over \tilde\omega_k-\epsilon_{k'}+
\epsilon_{-}(k')+i\Delta}+
{1 \over \tilde\omega_k+\epsilon_{k'}-
\epsilon_{+}(k')+i\Delta}]+ \nonumber \\
2 \sum_{k',k''} {T_{k'}^2 <\hat n_{k'}>\over
\tilde\omega_k-\epsilon_{k'}+
\epsilon_{-}(k')+i\Delta}\times
{1 \over
\tilde\omega_k+\epsilon_{k''}-\epsilon_{k'}}\times
{T_{k''}^2 <\hat n_{k''}> \over
\tilde\omega_k+\epsilon_{k''}-
\epsilon_{+}(k'')+i\Delta}+ O(T^8)
\label{sigma2-4}
\end{eqnarray}
 where
$\epsilon_{\pm}(\epsilon_{k'})=
\epsilon_0-{\Delta \over \pi}
Ln({\epsilon_{k'} \pm \tilde\omega_k \over D})$.
We also get
$\Sigma_3(\tilde\omega_k)=\Sigma_2(\tilde\omega_k)-
\Sigma_1(\tilde\omega_k)$, 
$\Sigma_4(\tilde\omega_k)=\Sigma_2(\tilde\omega_k)+
{\Sigma_3(\tilde\omega_k)^2 \over
\tilde \omega_k-\Sigma_2(\tilde\omega_k)}$, as before. 
These higher order self-energies give
rise to the first non-zero contribution to the 
Kondo resonance decoherence, due to the cross terms
between the right and left leads, as has already 
been pointed out in \cite{WM1}.
Eqs.(\ref{sigma1-4}) and (\ref{sigma2-4})
represents the first terms of an expansion for
$\Sigma_1$ or $\Sigma_2$, using 
$(T_k/\epsilon_0)^2$ as the expansion parameter, and
allows us to analyze the Kondo
resonance decoherence as a function of an external
bias. For $V=0$, we again find 
$\Sigma_1(\tilde\omega_k \rightarrow 0)\rightarrow 0$
and 
$\Sigma_4(\tilde\omega_k \rightarrow 0)\rightarrow 0$ 
 but for $V\neq 0$ 
these self-energies develop
non-negligible imaginary parts for $\tilde\omega_k
\rightarrow 0$.
In the bias regime,
defined by 
$V<<|\tilde \epsilon_0|<
\Delta$, we find 

\begin{equation}
\Sigma_1(\tilde\omega_k \rightarrow 0; V)
\simeq
-i({\Delta \over \tilde\epsilon_0})^2
{V \over 2\pi} (1-2{\Delta \over
\pi |\tilde\epsilon_0|}Ln{V \over |\tilde\epsilon_0|})
\equiv -i\gamma
\label{sigmaV}
\end{equation}
and $\Sigma_2 \simeq -i2\Delta$. In eq.(\ref{sigmaV}), 
$\tilde\epsilon_0$ is a renormalized level given
approximately by
$\tilde\epsilon_0 \simeq \epsilon_0+{\Delta \over \pi}
Ln{\pi D \over \Delta}$ .
Fig.4 shows the evolution of the  Kondo-like
resonance spectral density  for the same parameters used in
Fig.1, but introducing different biases on the QD.
Here we have used eq.(\ref{V4}) and the self-energies
to $T^6$-order.
Our results are compared with the NCA-calculations of
Ref. \cite{Pli}, where available. The agreement between
both approaches, for $\omega>0$, is remarkable. 
The difference
between both cases can be attributed to the 
bandwidth  used in \cite{Pli}, which is ten times smaller
than ours. For $\omega<0$, however, the agreement is
only reasonable at low bias (like in the equilibrium
case shown in Fig.1), which improves as $V$ increases.

We have also calculated the QD-conductance which, in the
symmetric configuration we are considering, is basically
proportional to the mean value of the spectral density
in the interval $(-V/2,V/2)$.  Fig.5
shows the conductance $G$ in units of the conductance
quantum $G_0=2e^2/h$,
as a function of $V/T_K$; a good fit
is obtained by the  expression:

\begin{equation}
G/G_0= 1/(1+0.75 Ln(1+({V \over 2T_K})^2)),
\label{conductance}
\end{equation}
for large values of $V/T_K$. 
These values also show  good agreement with the
NCA-calculations of \cite{RPT} for $V>2 T_K$, where our
solution can be taken with great confidence, 
although in this case, 
$G/G_0$ behaves like $1/Ln^2(V/T_K)$ for very high
voltages, $Ln(V/T_K)>>1$.
  
In addition, we have also addressed the important issue of strong
versus weak coupling regimes, as a function of the bias.
It has been argued \cite{Col1} that the strong-coupling
regime might subsist, even for high bias ($V>>T_K$), if
the decoherence rate $\gamma$ is smaller than
$T^{*}=-V/2+\sqrt{T_K^2+(V/2)^2}$.  With $\gamma$ given
by eq.(\ref{sigmaV}), the strong-coupling regime is
obtained for
$V\leq \pi^{1/2}{|\tilde\epsilon_0| \over \Delta}T_K$.
This is a very stringent condition; for instance, for
the values of the parameters in Figs.1a and 2, we find
strong coupling for $V\leq 3.5 T_K$, for which the
conductance is larger than $0.4G_0$. We conclude that,
when subjected to an external bias $V>T_K$, the QD-Kondo resonance
develops a two-peaked structure at $\pm V/2$ for
$V>2T_K$;
the system, however, evolves quickly as a function of
the voltage, moving from a strong-coupling regime with
the spin impurity screened by the electrons of the
leads, to a weak-coupling regime with that spin
unscreened.

\section{Conclusions}
The EOM method provides an
accurate approximation
to the Kondo problem in and out of equilibrium
situations, the
only requirement being to include all the Green
functions in
a strict expansion in $T^2$. 
Our results to $T^4$-order in equilibrium 
provide a good description of the Kondo
peak in the Kondo regime and an excellent
aproximation to (numerically) exact results in the
mixed-valence and empty-orbital regimes. 
In the nonequilibrium regime, it is
necessary to carry the method to  $O(T^6)$ 
 to find the Kondo
resonance decoherence, since it involves the
coupling between the right and left electrodes through
the QD. In particular, in our approach, 
the decoherence rate is
found analytically to be nearly proportional to $V$,  
which indicates that the system is driven readily
 to the weak couping
regime if
$V\geq \
\pi^{1/2}{|\tilde\epsilon_0| \over \Delta}T_K$,
with the spin impurity unquenched by the electrons of
the leads. 

\section{Acknowledgements}
We are indebted to E.A. Anda and E.C. Goldberg for
many
fruitful discussions,
to S. Davison for a critical reading of the manuscript 
and to T. Costi and P. Nordlander
for providing us with their data. This work has been
funded by the Spanish Comisi\'on Interministerial de
Ciencia y Tecnolog\'{\i}a under contracts MAT-2001-0665
and BFM-2001-0150.

\begin{figure}
\includegraphics[width=14cm]{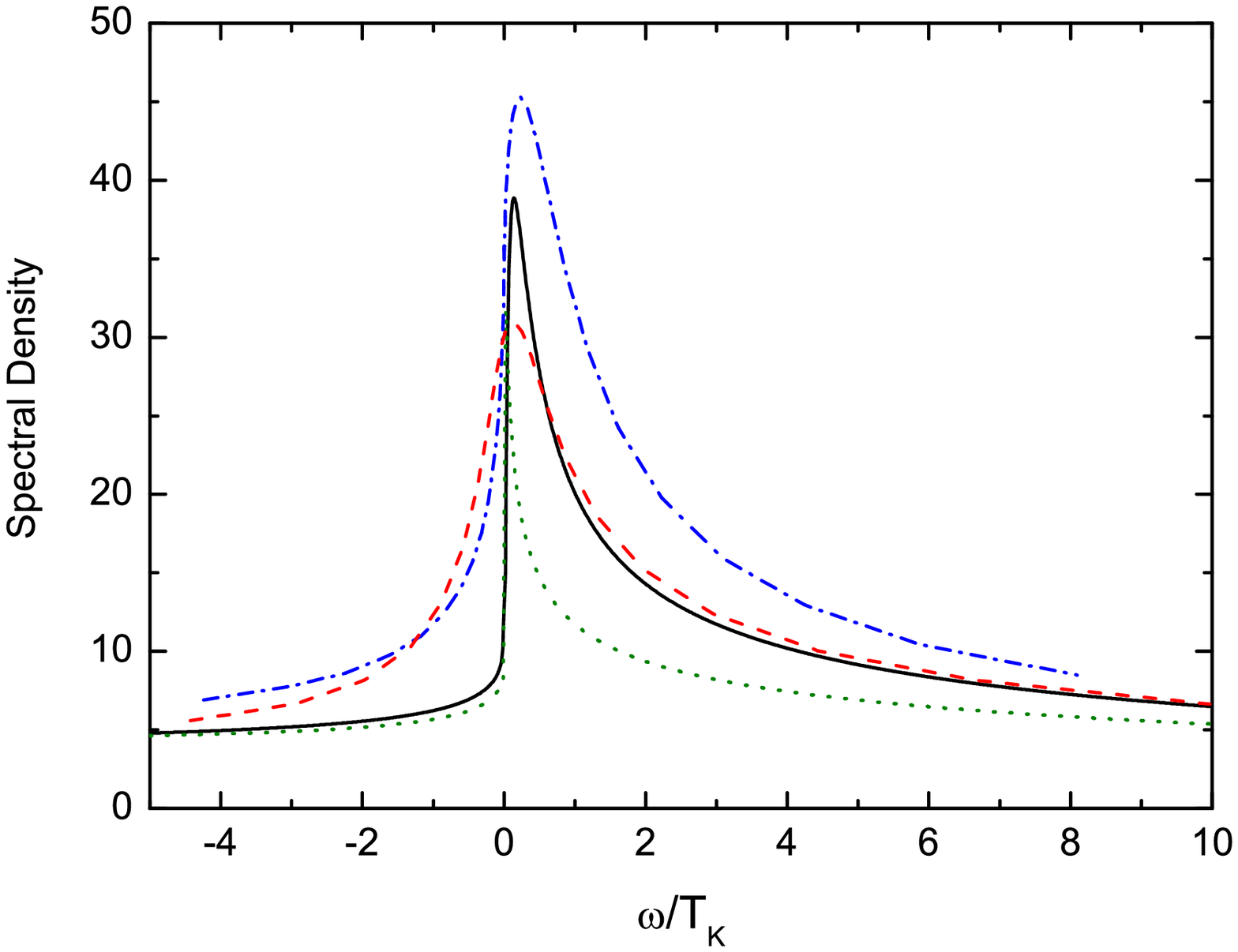}
\caption{\label{fig:fig1} The spectral density in equilibrium 
as a function of $w/T_K$
for $D=1$ and $\Delta=0.01$ is plotted for  
the Kondo regime: $\epsilon_0=-4\Delta$,
$T_K/\Delta=1.87\times 10^{-2}$.
 Dotted line: $T^2$-EOM; full line:
$T^4$-EOM; dashed line: NRG from [21]: dashed-dotted
line: NCA from [21]($T=10^{-6}D$).}
\end{figure}

\begin{figure}
\includegraphics[width=14cm]{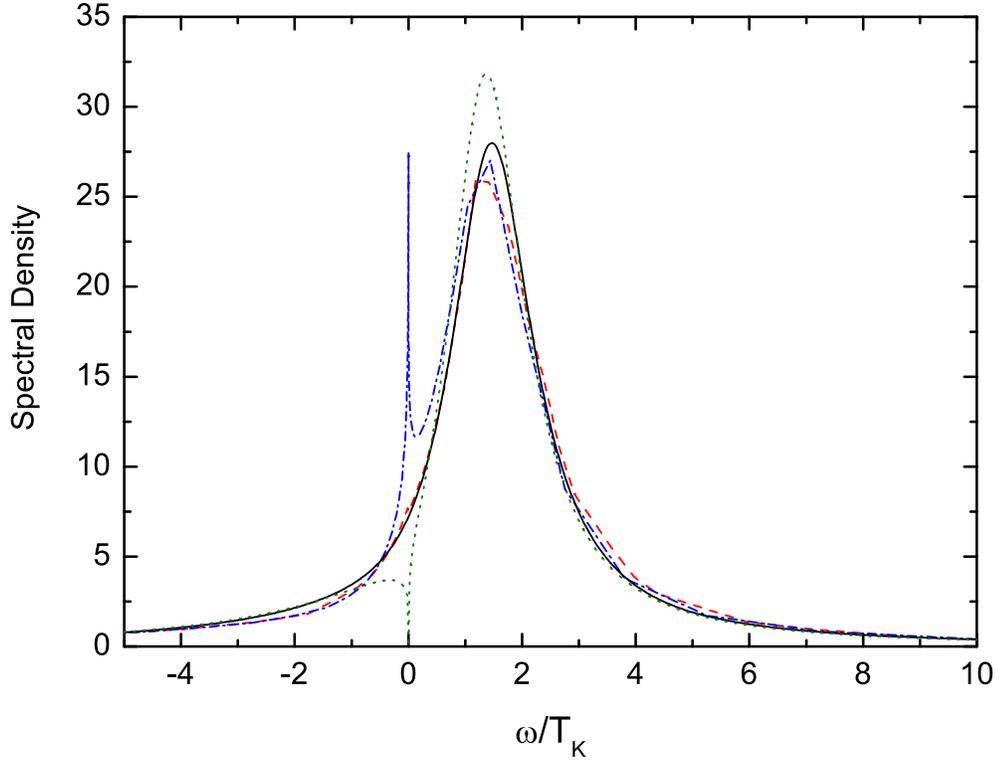}
\caption{\label{fig:fig2} The spectral density in equilibrium 
as a function of $w/T_K$
for $D=1$ and $\Delta=0.01$ is plotted for  
the mixed-valence regime $\epsilon_0=0$,
$T_K/\Delta=1$. Dotted line: $T^2$-EOM; full line:
$T^4$-EOM; dashed line: NRG from [21]: dashed-dotted
line: NCA from [21]($T=10^{-6}D$).}
\end{figure}

\begin{figure}
\includegraphics[width=14cm]{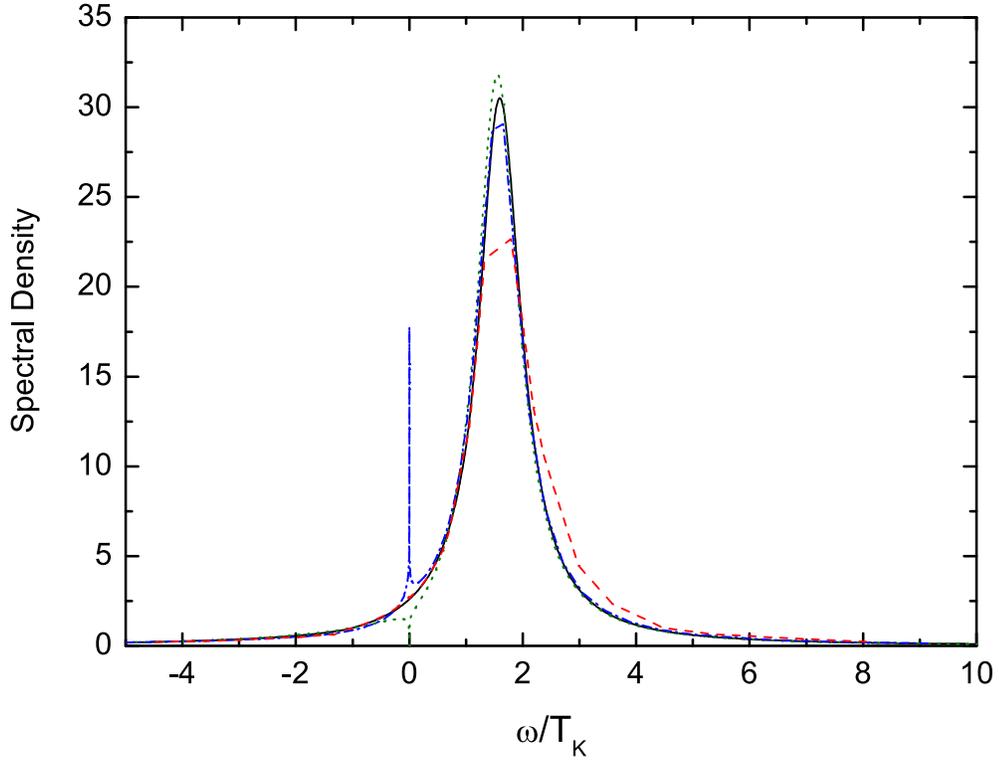}
\caption{\label{fig:fig3} The spectral density in equilibrium 
as a function of $w/T_K$
for $D=1$ and $\Delta=0.01$ is plotted for  
the empty-orbital regime $\epsilon_0=2\Delta$,
$T_K/\epsilon_0=1$. Dotted line: $T^2$-EOM; full line:
$T^4$-EOM; dashed line: NRG from [21]: dashed-dotted
line: NCA from [21]($T=10^{-6}D$).}
\end{figure}

\begin{figure}
\includegraphics[width=12cm]{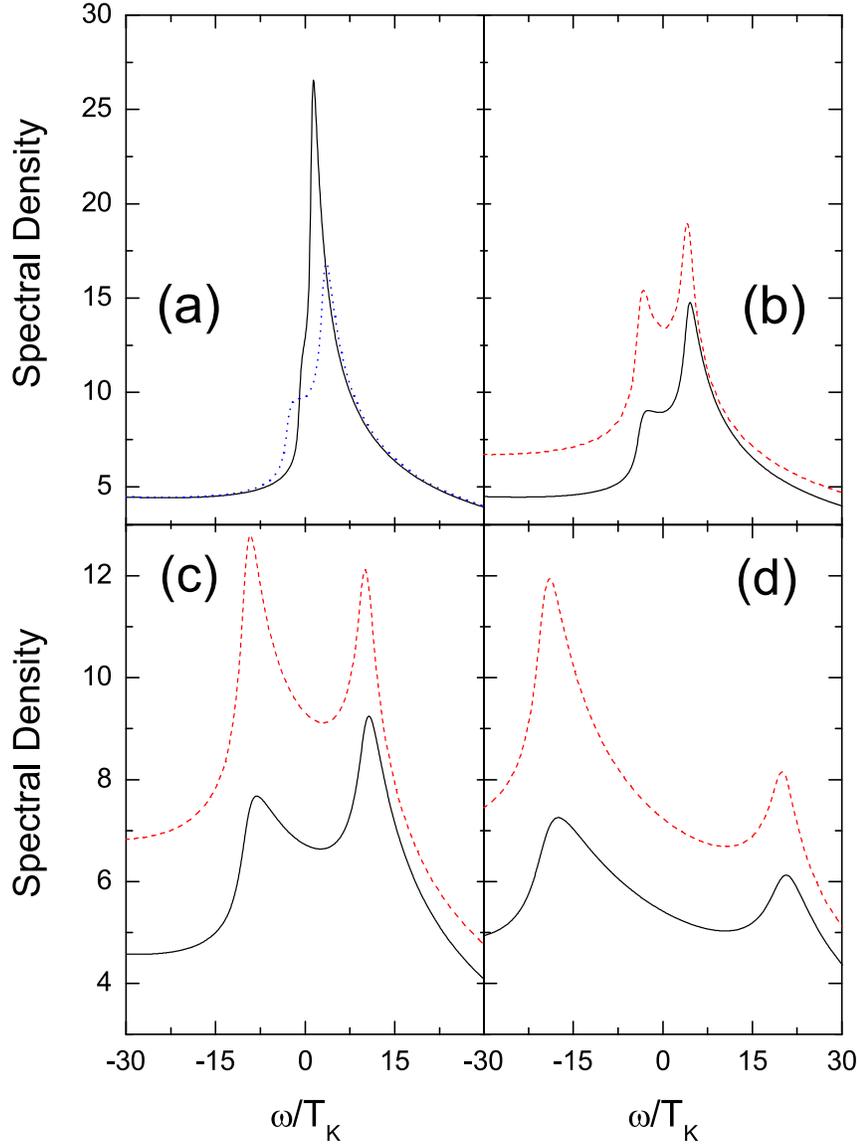}
\caption{\label{fig:fig4} The spectral density in the 
nonequilibrium regime as a function of $w/T_K$, 
calculated with our EOM method for the same
parameters as in Fig.1 is
shown for: a) $V=2T_K$ (continuous line) 
and $V=6T_K$
(dotted line). In Figs.4b, 4c and 4d our results
(continuous lines) are
compared with the
NCA results of Ref.[10] (dashed lines) for:
b) $V=8T_K$, c) $V=20T_K$, d) $V=40T_K$.}
\end{figure}

\begin{figure}
\includegraphics[width=12cm]{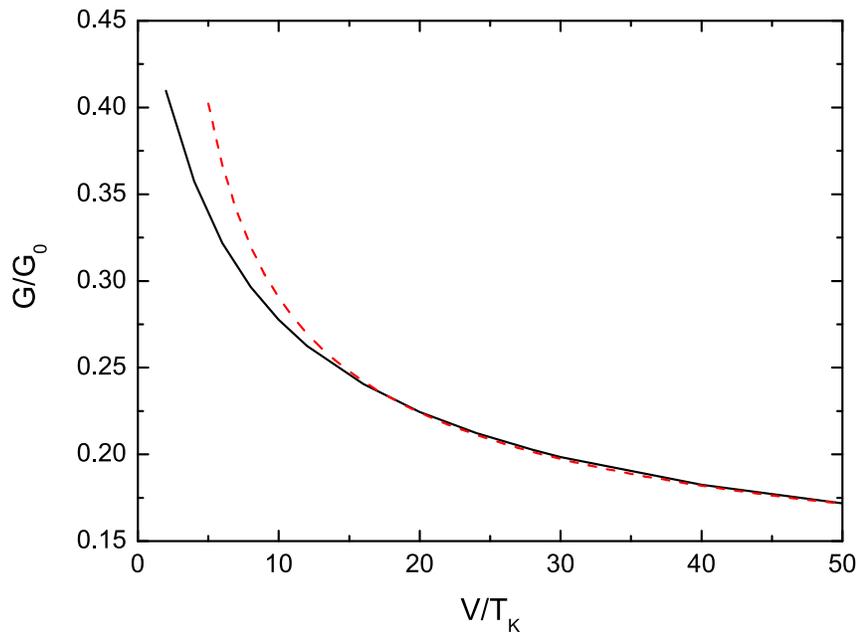}
\caption{\label{fig:fig5} 
The conductance (in units
of the quantum of conductance) calculated for the same
parameters as in Figs.1 and 4 (continuous line) 
is compared with the fitting 
function of eq.(\ref{conductance}) (dashed line).}
\end{figure}

\end{document}